\documentclass[aps,prl,twocolumn,showpacs,groupedaddress]{revtex4}

\usepackage{graphicx}
\begin{document}


\title{Increase of the mean inner Coulomb potential in Au clusters induced by surface tension and
its implication for electron scattering}


\author{Radian Popescu}
\email[Electronic address: ]{popescu@lem.uni-karlsruhe.de}
\author{Erich M\"uller}
\author{Matthias Wanner}
\altaffiliation{present adress: Forschungsinstitut f\"ur Pigmente und Lacke e.V., D-70569 Stuttgart, Germany}
\author{Dagmar Gerthsen}
\affiliation{Laboratorium f\"ur Elektronenmikroskopie and Center for
  Functional Nanostructures (CFN), Universit\"at Karlsruhe, D-76128 Karlsruhe, Germany}
\author{Marco Schowalter}
\author{Andreas Rosenauer}
\affiliation{Institut f\"ur Festk\"orperphysik, Universit\"at Bremen, D-28359 Bremen, Germany}
\author{Artur B\"ottcher}
\author{Daniel L\"offler}
\author{Patrick Weis}
\affiliation{Institut f\"ur Physikalische Chemie and CFN, Universit\"at Karlsruhe, D-76128 Karlsruhe, Germany}


\date{\today}

\begin{abstract}
  Electron holography in a transmission electron microscope was applied to measure the 
  phase shift $\Delta \varphi$ induced by Au 
  clusters as a function of the cluster size. Large $\Delta \varphi$ observed for small 
  Au clusters cannot be described by the well-known equation
  $\Delta \varphi=C_E V_0 t$ ($C_E$: interaction constant, $V_0$: mean inner Coulomb
  potential (MIP) of bulk gold, $t$: cluster thickness). The rapid increase of the Au MIP
  with decreasing cluster size derived from $\Delta \varphi$, 
  can be explained by the compressive strain of surface atoms in the cluster.
\end{abstract}


\maketitle

Au clusters are considered as prototype material for nano-scaled electronic
devices and biosensors \cite{Andres96}. Moreover, Au clusters exhibit an exceptional
catalytic activity \cite{Haruta97}. All these potential applications have motivated numerous
studies regarding the properties of Au nano-clusters.
{\it One fundamental material property is the mean inner Coulomb
potential (MIP), which plays an important role for the quantitative evaluation
of experimental data obtained from electron scattering techniques, 
e.g. transmission electron microscopy (TEM) and electron holography (EH).
The MIP is the volume-averaged electrostatic part of the crystal
potential, which can be expressed \cite{Bethe28,Keeffe94} by}
\begin{equation}
\label{eq:MIP_def}
V_0=\frac{h^2}{2\pi m e \Omega}\sum_i n_i f_i^{el}(0)\quad\text{,}
\end{equation}
with Planck's constant $h$, the electron mass and charge $m$ and $e$, the unit
cell volume $\Omega$ and the occupation number $n_i$ for the atomic
species $i$ within the unit cell. {\it The important property in Eq.(1) is the atomic
scattering factor $f_i^{el}(0)$ \cite{Doyle68} which correlates 
the MIP with the amplitude of the electron wave scattered in forward direction. The MIP can be
determined by off-axis EH under kinematical diffraction conditions according to the relation
$\Delta \varphi=C_E V_0 t$ ($C_E$: interaction constant) by measuring the phase shift  $\Delta\varphi$
between the electron wave passing through the sample with a known thickness $t$ and a vacuum 
reference wave \cite{Reimer}.} On the other hand, local TEM sample thicknesses can be determined 
by EH, if precise values of the MIP are known, but thus far only MIP values for few materials with limited
accuracy are available \cite{Goswami82,Spence93,Kruse03,Gajdardziska,Schowalter06}.
For instance, experimental values for the MIP of Au between 16.8 and 30.2 V were
reported \cite{Goswami82}, whereas calculations yield values of 25.0
to 35.9 V \cite{Goswami82,Gajdardziska,Schowalter06}. Moreover, a strong
increase of the Au MIP up to 45 V was reported for Au clusters deposited on
$TiO_2$ powder with decreasing cluster size \cite{Ichikawa03}. Recently,
{\it effective} carbon MIP values up to 65 V were reported for ultra-thin amorphous carbon
(a-C) films compared to a bulk value of 9 V \cite{Wanner06}. This indicates that the MIP increase for
nano-scaled objects could be a general phenomenon. {\it In this study, we
applied EH to determine 1) the MIP of bulk Au, which corresponds to the MIP of Au atoms in the 
cluster core and 2) the contribution of surface atoms to the overall MIP of Au clusters to elucidate 
the physical origin of its increase.}

Samples were prepared by low-energy-beam cluster deposition of $Au_n$
clusters with 10$\leq$$n$$\leq$20 atoms on commercial a-C
substrates, $\approx$10 nm thick. Due to the storage of the sample, a coarsening 
of the particle sizes occurs, which leads to Au clusters with diameters D between 0.8$\leq$$D$$\leq$8.0 nm.
Off-axis transmission EH was carried out in a 200 keV Philips CM200 FEG/ST
electron microscope equipped with a  M\"ollenstedt-D\"uker biprism in the
selected-area aperture holder. Holograms with an interference fringe distance
of (0.16$\pm$0.05) nm and a corresponding resolution $\Delta D$=(0.32$\pm$0.10) nm were
recorded using a 2048$\times$2048 pixel CCD camera. The $\Delta \varphi$ was reconstructed
from the hologram sideband \cite{Lehmann02} by using the DALI program package extended for
hologram reconstruction \cite{Rosenauer96}. For our microscope a constant
$C_E$=7.29$\cdot$10$^{\text{6}}$ rad$\cdot$(Vm)$^{-1}$ was determined according to Ref. \cite{Kruse03}.

\begin{figure}
 \includegraphics[width=7cm]{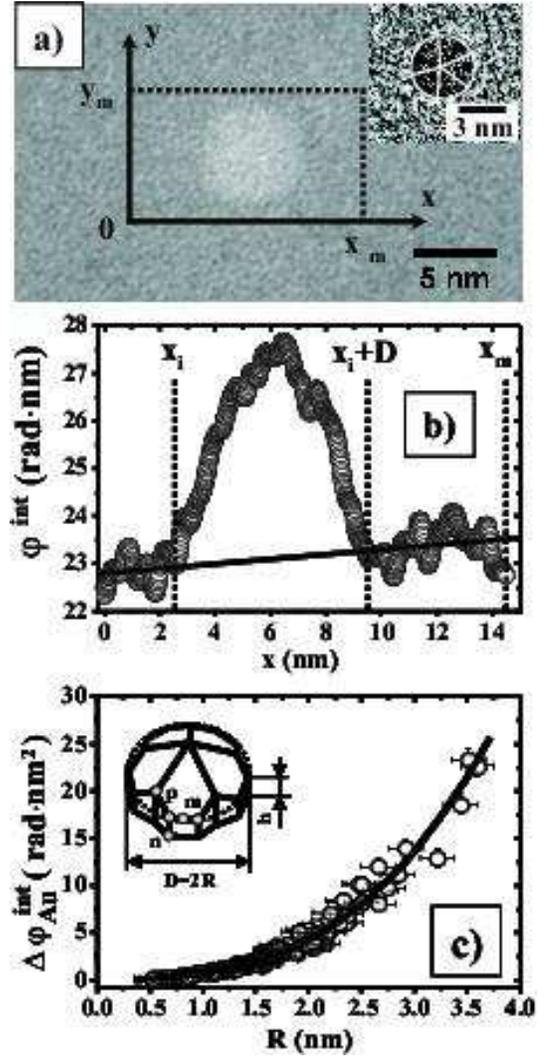}
 \caption{\label{fig:fig1} a) Gray-scale coded (256 gray levels between
   black (no phase shift) and white (phase shift equal to $2\pi$ rad))
   reconstructed phase shift obtained for an Au cluster. The
   frame represents the integration domain (see text). 
   The insert shows a typical HRTEM image of an Au cluster with M-Dh structure;
   b) $\bigcirc$: the experimental phase shift integrated
   along the y-direction. The solid line indicates the linearly interpolated
   background. c) Experimental ($\bigcirc$) and
   calculated (solid line) integrated phase shift of Au clusters 
   $\Delta\varphi^{int}_{Au}$($R$). The insert shows
   an M-Dh Au cluster with our experimental shape according to \cite{Cleveland97}.}
\end{figure}

Fig. \ref{fig:fig1}a) shows the reconstructed phase shift for an Au
cluster with D=6.9 nm. The total phase shift of the electron wave in the
bright region of Fig. \ref{fig:fig1}a) is given by the phase shift induced by
the cluster and the supporting a-C film, whereas the gray background represents 
the phase shift due to the a-C film only. We use
the following procedure to extract the integrated phase shift induced by the
Au cluster $\Delta \varphi_{Au}^{int}$: first, the integration of the
phase shift along the y-direction of the integration domain
(black frame in Fig. \ref{fig:fig1}a)) was performed according to:
$\varphi^{int}(x)=\int_0^{y_m}\varphi(x,y)dy$ (see Fig. \ref{fig:fig1}b)).
To eliminate the substrate contribution $\varphi_{sub}^{int}(x)$, the background is 
linearly interpolated (straight line in Fig. \ref{fig:fig1}b)) on both sides of the 
cluster. Finally, $\Delta \varphi_{Au}^{int}$ is obtained by integration along the x-direction:
$\Delta\varphi^{int}_{Au}=\int_{x_i}^{x_i+D}\left(\varphi^{int}(x)-\varphi^{int}_{sub}(x)\right)dx$.
In Fig. \ref{fig:fig1}c) $\Delta \varphi_{Au}^{int}$ of 123
Au clusters is plotted versus the radius of cluster projection $R$.

To analyze the experimental $\Delta \varphi_{Au}^{int}$ we propose
a new expression for the MIP of Au clusters which distinguishes between
{\it surface} and {\it core} atoms. Generally, the atoms in a cluster are compressed due
to surface tension. Previously observed compressive strain in metallic
clusters (Au, Cu, Ni, Pt) was attributed to the sole contraction of the
atoms at the cluster surface \cite{Cleveland97,Apai79,Moraveck79,Kluth04},
which is essential for the explanation of the lattice vibrations in small
particles \cite{Tamura82}. Considering that the strain is confined entirely to
the cluster surface, {\it surface} atoms are under uniform compressive strain,
whereas {\it core} atoms are unstrained. The strain $\varepsilon$ leads to a decrease
of the atomic volume for {\it surface} atoms to $(1+\varepsilon)^3\Omega_{at}$, where $\Omega_{at}$ is the volume of
unstrained {\it core} atoms, which is identical to the volume of bulk atoms.
The inverse dependence of the MIP on the volume ($V_0 \sim 1/\Omega$, see
Eq.(\ref{eq:MIP_def})) requires accordingly the distinction between the {\it surface} and {\it core}
atoms with different atomic volumes. The MIP of Au in Au clusters $V_0^{cl}$ can then be expressed by
\begin{equation}
\label{eq:MIP_surfstrain}
V_0^{cl}=\left(1-\frac{N_S}{N_T}\right)V_0+\frac{N_S}{N_T}\frac{V_0}{(1+\varepsilon)^3}\quad\text{,}
\end{equation}
with the number of {\it surface} atoms $N_S$ and the total number of atoms
within the cluster $N_T$. $V_0$ is the MIP value of {\it core} atoms, which is equal to
the MIP of the bulk material. As expected, $V_0^{cl}$ approaches $V_0$ for bulk material, if
$N_S/N_T$$\to$$0$ and $\varepsilon$$\cong$$-0.02$. However, large differences can be anticipated
for nano-scaled objects, which are characterised by increased $N_S/N_T$ ratios and significant
strain $\varepsilon$. A theoretical $\Delta \varphi_{Au}^{int}$ can then be calculated on the basis of
Eq.(\ref{eq:MIP_surfstrain}) as
\begin{equation}
\Delta \varphi^{int}_{Au}=C_E V_0^{cl}\int \! \int_{\Sigma} t(x,y)=C_E V_0^{cl} \Omega_{cl}\text{,}\label{eq:PhaseShift_theo}
\end{equation}
where $t(x,y)$ is the cluster thickness at a given position within the area of
the cluster projection $\Sigma$ on the (x,y) hologram (image) plane and
$\Omega_{cl}$ denotes the geometrical cluster volume.

To interpret the measured $\Delta\varphi^{int}_{Au}$ using Eq.(\ref{eq:PhaseShift_theo}), the cluster
shape and structure needs to be known.  According to Fig. \ref{fig:fig1}a) the area of the Au-cluster
projection corresponds in a good approximation to disks with radii $R$
suggesting that the clusters could be spheres. To verify this assumption,
high-resolution TEM (HRTEM) images of Au clusters with $0.6$$\leq$$R$$\leq$$3.4$ nm 
were recorded at normal illumination ($0^\circ$) and after tilting the samples
by $\pm22^\circ$. The projected cluster area increases by about 5\% in the tilted
position, which excludes spherical clusters. Vertical facets are required
with a ratio $h/R$$\cong$$0.4$ of the vertical facet height ($h$) with respect to the
apparent radius $R$ of the cluster projection.

Calculations indicate that clusters with a pentagonal decahedral (Dh)
structure variant called Marks-Dh (M-Dh) \cite{Marks84} and the
face-centered-cubic ({\it fcc}) truncated octahedral (TO) clusters are energetically
the most stable structures for $Au_N$ clusters ($50$$\leq$$N$$\leq$$5000$ atoms)
\cite{Cleveland97,Wetten96,Cleveland97_2}. A growth study of small Au particles (size 4-25 nm) produced by gas
evaporation in flowing Ar has shown that, after condensation of the clusters
on a substrate, the growth of Au particles with icosahedral (Ih) or Dh
structures is favored with respect to {\it fcc} ones \cite{Renou81}. Accordingly, 
the Au clusters may have Ih or M-Dh structures. But except for the smallest Au
clusters, it was shown that the Ih structures are energetically noncompetitive
as compared with the M-Dh ones \cite{Cleveland97,Cleveland91}. We therefore
assume the M-Dh morphology for Au clusters in our experimental size range. This 
assumption is confirmed by HRTEM images showing the typical 
fivefold symmetry (insert in Fig. \ref{fig:fig1}a)). The M-Dh Au clusters are
characterized by $h$=0.4$R$, derived from the tilt experiments, and an apparent
diameter of $D$=2$R$ (see the insert in Fig. \ref{fig:fig1}c)). The latter relation
is assumed to be valid because the cluster projections are almost circular.

To evaluate Eqns.(\ref{eq:MIP_surfstrain}) and (\ref{eq:PhaseShift_theo}), $N_S/N_T(R)$
and $\varepsilon(R)$ for Au clusters with our shape are required. {\it For the theoretical
estimation of these properties we distinguish between: a) ideal M-Dh Au clusters
with closed-shell structures \cite{comment1} and corresponding ideal radii $R_{id}$ and
b) M-Dh Au clusters without closed-shell structures and intermediate radii $R$.}
We calculate first $N_S/N_T$ and $\varepsilon$ for the ideal clusters.
The total number of Au atoms in ideal M-Dh clusters, i.e. the "magic numbers",
$N_{T\_id}$ can be described by the number of atoms $n$, $m$ and 
$p$ \cite{Cleveland97,Urban98} as shown in the insert in Fig. \ref{fig:fig1}c). We use 
only $p$=2 to approximate the observed circular 
cluster cross section. For a given $m$, $n$ is calculated on the basis of the experimental 
ratio $h/R$. Then, the difference of $N_{T\_id}$ in clusters with {\it(i+1)}
and {\it (i)} closed shells, $N_{T\_id}^{(i+1)}-N_{T\_id}^{(i)}=N_{S\_id}^{(i+1)}$, 
corresponds to the number of {\it surface} atoms of an ideal cluster with {\it(i+1)} shells.
\begin{figure}
 \includegraphics[width=7cm]{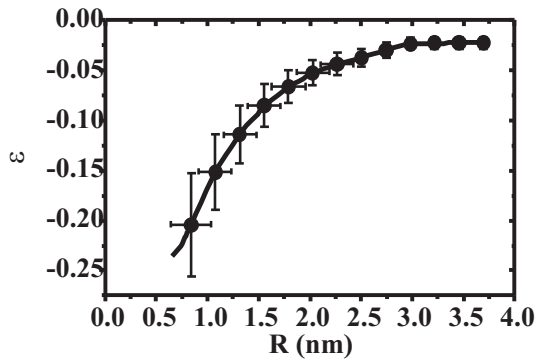}
 \caption{\label{fig:fig2} Uniform compressive strain of {\it surface} atoms
   $\varepsilon$($R$) for M-Dh Au clusters ($\bullet$: ideal M-Dh clusters).}
\end{figure}

The following considerations yield estimates for the strain $\varepsilon$ 
and cluster radii $R_{id}$ which depend themselves on $\varepsilon$.
Implying that $\varepsilon$ is confined completely to the cluster
surface, an ideal M-Dh Au cluster with {\it (i+1)} closed shells and radius $R_{id}^{(i+1)}$ is formed by
a) an unstrained M-Dh cluster core with {\it(i)}-closed shells and radius $R_{us}^{(i)}$ 
consisting of $N_{T\_id}^{(i)}$ {\it core} atoms and
b) the {\it(i+1)}-closed shell, formed by $N_{S\_id}^{(i+1)}$ 
{\it surface} atoms with radii $R_{Au\_S}$$<$$R_{Au}$.
We assume that the unstrained M-Dh cluster core consists of
unstrained {\it core} Au atoms with a diameter $d_{Au}$=$2R_{Au}$=$0.288$ nm,
which is given by the minimum bond length between Au atoms in bulk gold.
The strain $\varepsilon^{(i+1)}$ is proportional
to the reduction of the minimum bond lengths between {\it surface} and {\it core} atoms as compared 
to $d_{Au}$: $\varepsilon^{(i+1)}=[(R_{Au}+R_{Au\_S})-2R_{Au} ]/2R_{Au}$. However,
a more useful expression for $\varepsilon^{(i+1)}$ is given by:
$\varepsilon(R_{id}^{(i+1)})\approx[(R_{id}^{(i+1)}-R_{us}^{(i)})-2R_{Au}]/2R_{Au}$.
$R_{us}^{(i)}$ can be estimated by constructing
unstrained M-Dh clusters with closed-shell structure using only geometrical
considerations and Au atoms with $R_{Au}$. To estimate $R_{id}^{(i+1)}$, we
define the atom density of clusters as $\rho(R_{id})$=$N_{T\_id}$/$\Omega_{cl}(R_{id})$. 
For M-Dh Au clusters with our experimental shape, an analytical expression
of $\Omega_{cl}(R_{id})$ can be derived. The lower boundary for $R_{id}$ (and the 
upper boundary for $\varepsilon$) is calculated considering that $\rho(R_{id})$ cannot 
be larger than the atom density of $fcc$ bulk Au $\rho_{bulk}$=58.9 atoms/nm$^3$ by solving
$N_{T\_id}^{(i+1)}/\Omega_{cl}(R_{id}^{(i+1)})$=$\rho_{bulk}$. The upper boundary
for $R_{id}$ (and the lower boundary for $\varepsilon$) is given by
$\rho_{cl}$=48.0 atoms/nm$^3$ obtained for a cluster with two closed shells from
{\it ab initio} density functional theory calculations, which
yields $\varepsilon=-15.9\%$. The atom density of large Au clusters with
$\geq$$10^6$ atoms is $\rho_{cl}$$\cong$$\rho_{bulk}$ and the strain
$\varepsilon$ in these clusters converges towards the strain of surface atoms
in bulk Au between -1 and -2\% for low-index metal surfaces
\cite{Jona88}. The strain $\varepsilon$($R_{id}$) in ideal M-Dh Au clusters approximated
by the average values on the basis of $\rho_{bulk}$ and $\rho_{cl}$ is plotted in 
Fig. \ref{fig:fig2}. The error bars represent strain values deduced from $\rho_{bulk}$ and $\rho_{cl}$.
We note that $R_{id}$ differences associated with maximum and minimum strain are
smaller than the spatial resolution in our experiment.

Values for $\varepsilon$ and $N_S/N_T$ for M-Dh Au clusters without closed shell
of surface atoms are obtained by interpolation between the properties of the ideal ones.
The MIP value of {\it core} Au atoms, identical to the MIP of bulk Au,
is then calculated by a least-square fit of the experimental
$\Delta \varphi_{Au}^{int}(R)$ and calculated phase shift on the basis of 
Eq.(\ref{eq:PhaseShift_theo}) with $V_0$ as the only free parameter (solid line in
Fig. \ref{fig:fig1}c)). We obtain $V_0$=$(32.2\pm3.6)$ V, which agrees well with
calculated MIP values of bulk Au \cite{Goswami82,Schowalter06} and reach, within the
error bar, the upper limit of the previous experimental MIP data
\cite{Goswami82}.

\begin{figure}
 \includegraphics[width=7cm]{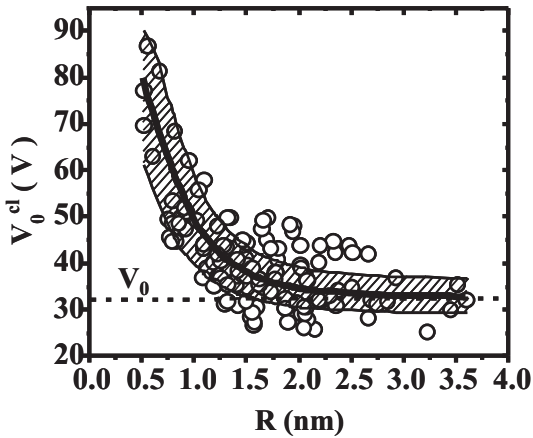}
 \caption{\label{fig:fig3}  Experimental ($\bigcirc$) and calculated (solid
   line) Au MIP values in M-Dh Au clusters $V_0^{cl}$($R$). The hatched region
   represents the maximum error of the $V_0^{cl}$ values
   induced by the upper and lower strain estimates.}
\end{figure}

The experimental MIP values of Au in Au clusters are estimated from 
the measured $\Delta\varphi_{Au}^{int}(R)$ by using
$V_0^{cl}(R)={\Delta \varphi_{Au}^{int}(R)}/{C_E \Omega_{cl}(R)}$ (see Fig. \ref{fig:fig3}). 
The solid line in Fig. \ref{fig:fig3} is calculated using Eq.(\ref{eq:MIP_surfstrain})
with $V_0$=$32.2$ V, which is the fitted MIP of {\it core} Au atoms.
Despite the scattering of $V_0^{cl}$ for 1.5$\leq$R$\leq$2.2 nm, 
Eq.(\ref{eq:MIP_surfstrain}) describes adequately the behavior of the MIP.
The description is particularly reasonable for small clusters with $R\approx0.5$ nm where
$V_0^{cl}$=85 V is obtained. Data points exceeding the error
limits indicated in Fig. \ref{fig:fig3} can be attributed to Au clusters with
different $h/R$ ratios, the estimation of the background phase shift or a
tilted position of the cluster on the a-C substrate. The errors associated with
different possible cluster structures (M-Dh or Ih) on the resulting MIP are smaller
than the error limits.

To explain the MIP increase in Au clusters and thin a-C films, only effects
of changes of the electronic structure, permanent electrical charging or
adsorbate molecules with a large electrical dipole have been
considered up to now \cite{Keeffe94,Ichikawa03,Wanner06}. Ichikawa et al.
\cite{Ichikawa03} attributed the increase of the Au MIP to charge transfer
from the Au clusters to the $TiO_2$ substrate. However, we exclude persistent 
charging of the clusters in our study, because the Au clusters are deposited 
on a-C substrate in electrical contact with the metallic sample holder.
{\it The presence of charged and neutral impurities 
has to be considered. The effect of neutral impurities is expected to be negligible 
due to the small amount of additional charge density. Charged impurities generate 
an electrostatic potential, which contributes to the measured phase shift. However,
the consisting of our data set indicates that this is not a serious problem.
We also add that deformations of atoms on twin planes in M-Dh clusters cannot be 
distinguished from compressed surface atoms, but this effect is estimated to be small 
compared to surface tension. Finally, effects of phonons and surface plasmons in clusters on the 
measured phase shift can be excluded due to the inelastic nature of the scattering process.}
In our case, the strong increase of the Au MIP
in Au clusters can be well explained by the compressive strain $\varepsilon$ of 
{\it surface} atoms. This effect is present but can not be measured in bulk samples, where
$N_S$/$N_T$ is negligible and $\varepsilon$ is small. In contrast, surface contributions 
dominate and relaxations are important in small clusters.

{\it In summary, we used EH to reconstruct the phase shift $\Delta \varphi$ induced by small Au clusters.
$\Delta \varphi$ is determined by the MIP, which can be modeled by distinguishing 
between {\it surface} and {\it core} atoms with different atomic volumes.
MIP values of 85 V are measured for Au clusters with $R$$\approx$$0.5$ nm, in 
contrast to the derived MIP value of Au {\it core} atoms of $V_0$=$(32.2\pm3.6)$ V 
in good agreement with previous calculations for bulk Au. Significant 
compression of surface atoms (reduction of bond length) yields a high surface potential 
contribution to the MIP of clusters. An essential additional ingredient of the model, which 
is generally applicable, is the increasing ratio $N_S$/$N_T$ in nano-scaled objects. The effect of 
the MIP increase in Au clusters and in many nano-scaled objects has important 
consequences for the quantification of electron scattering data in general because the amplitude 
of the electron wave scattered in forward direction can be much higher than expected from atomic 
scattering factors given e.g. in \cite{Doyle68}. Modified scattering factors should 
be considered for the evaluation of images and diffraction patterns of nano-scaled objects from 
techniques based on electron scattering.}

\begin{acknowledgments}
We thank M.M. Kappes (Institut f\"ur Physikalische Chemie,
Universit\"at Karlsruhe) and H.L. Meyerheim (Max-Planck Institut f\"ur
Mikrostrukturphysik, Halle) for fruitful discussions. This work has been
performed within the project C4 of the DFG Research Center for Functional
Nanostructures (CFN). It has been further supported by a grant from the
Ministry of Science, Research and the Arts of Baden-W\"urttemberg (Az:
7713.14-300).
\end{acknowledgments}

\bibliography{apssamp}

\begin{thebibliography}{28}
\bibitem[{com()}]{comment1}
\bibinfo{note}{Closed-shell cluster structures denote clusters with closed
  atomic shells which consist of a defined numbers of atoms - "magic numbers"-
  determined only by the geometric arrangement. Every additional closed shell
  represents a new layer of atoms which covers the whole cluster surface.}
\end{thebibliography}

\end{document}